\documentclass[twocolumn,showpacs,aps,superscriptaddress]{revtex4-1}
\usepackage{amsmath}
\usepackage{bm}
\usepackage{graphicx}
\usepackage{dcolumn}
\usepackage{float}
\usepackage{graphicx}
\usepackage{epstopdf}
\usepackage{epsfig}

\usepackage[colorinlistoftodos,prependcaption]{todonotes}

\begin{document}
\title{Spin dynamics of the ordered dipolar-octupolar pseudospin-$^1/_2$ pyrochlore Nd$_2$Zr$_2$O$_7$ probed by muon spin relaxation}

\author{J. Xu}
\altaffiliation{jianhui.xu@helmholtz-berlin.de}
\affiliation{\mbox{Helmholtz-Zentrum Berlin f\"{u}r Materialien und Energie GmbH, Hahn-Meitner Platz 1, D-14109 Berlin, Germany}}
\affiliation{\mbox{Institut f\"{u}r Festk\"{o}rperphysik, Technische Universit\"{a}t Berlin, Hardenbergstra$\beta$e 36, D-10623 Berlin, Germany}}
\author{C. Balz}
\affiliation{\mbox{Helmholtz-Zentrum Berlin f\"{u}r Materialien und Energie GmbH, Hahn-Meitner Platz 1, D-14109 Berlin, Germany}}
\author{C. Baines}
\affiliation{\mbox{Laboratory for Muon-Spin Spectroscopy, Paul Scherrer Institute, CH-5232 Villigen-PSI, Switzerland}}
\author{H. Luetkens}
\affiliation{\mbox{Laboratory for Muon-Spin Spectroscopy, Paul Scherrer Institute, CH-5232 Villigen-PSI, Switzerland}}
\author{B. Lake}
\altaffiliation{bella.lake@helmholtz-berlin.de}
\affiliation{\mbox{Helmholtz-Zentrum Berlin f\"{u}r Materialien und Energie GmbH, Hahn-Meitner Platz 1, D-14109 Berlin, Germany}}
\affiliation{\mbox{Institut f\"{u}r Festk\"{o}rperphysik, Technische Universit\"{a}t Berlin, Hardenbergstra$\beta$e 36, D-10623 Berlin, Germany}}
\date{\today}

\begin{abstract}
We present a muon spin relaxation study on the Ising pyrochlore Nd$_2$Zr$_2$O$_7$ which develops an "all-in-all-out" magnetic order below 0.4~K. At 20~mK far below the ordering transition temperature, the zero-field muon spin relaxation spectra show no static features and can be well described by a dynamical Gaussian-broadened Gaussian Kubo-Toyabe function indicating strong fluctuations of the ordered state. The spectra of the paramagnetic state (below 4.2~K) reveal anomalously slow paramagnetic spin dynamics and show only small difference with the spectra of the ordered state. We find that the fluctuation rate decreases with decreasing temperature and becomes nearly temperature independent below the transition temperature indicating persistent slow spin dynamics in the ground state. The field distribution width shows a small but sudden increase at the transition temperature and then becomes almost constant. The spectra in applied longitudinal fields are well fitted by the conventional dynamical Gaussian Kubo-Toyabe function, which further supports the dynamical nature of the ground state. The fluctuation rate shows a peak as a function of external field which is associated with a field-induced spin-flip transition. The strong dynamics in the ordered state are attributed to the transverse coupling of the Ising spins introduced by the multipole interactions.

\end{abstract}


\maketitle

\section{\label{Intro} INTRODUCTION}

In the field of geometrically frustrated magnetism, the rare-earth pyrochlores have been studied intensively for the last decades and many exotic states and emergent excitations have been found, such as the dipolar spin ice state and monopole-like excitations in  $R_2$Ti$_2$O$_7$ ($R$=Dy, Ho), the spin liquid phase in Tb$_2$Ti$_2$O$_7$ and the ordered state induced by quantum disorder in Er$_2B_2$O$_7$ ($B$=Ti, Sn) \cite{book,rev2010,castelnovo2008,morris2009,fennell2009}. The spin dynamics of these systems also show many interesting behaviors and have not been fully understood. Typical examples are the persistent spin dynamics or the coexistence of the spin dynamics and long-range order far below the transition temperature in Er$_2$Ti$_2$O$_7$, Tb$_2$Sn$_2$O$_7$ and Yb$_2$Sn$_2$O$_7$ and the anomalously slow paramagnetic spin dynamics in Yb$_2$Ti$_2$O$_7$ and Yb$_2$Sn$_2$O$_7$ \cite{lago2005,dalm2006,bert2006,yaouanc2013,hodges2002,mais2015}. For the classical dipolar spin ice phase, the dynamics were largely suppressed by the approximate Ising interactions and the spins tend to freeze at low temperatures \cite{rev2010,rau2015}. Recently, the concept of quantum spin ice (QSI) was introduced where a weaker-Ising interaction (containing additional pronounced transverse terms) enhances quantum fluctuations which allows coherent propagation of the magnetic monopoles with a linear dispersion similar to photons \cite{hermele2004, onoda2010, onoda2011, ross2011, savary2012,rau2015,kimura2013}.

Pyrochlores with Pr, Nd and Yb are proposed to be candidates for quantum spin ice where significant quantum fluctuations can be expected \cite{onoda2010,onoda2011,ross2011}. The microscopic mechanism for the quantum dynamics was established based on the effective pseudospin-1/2 model because the interacting rare-earth moments can be described by a crystal field ground state doublet at low temperatures \cite{ross2011,onoda2010,onoda2011}. For Nd pyrochlores, the crystal field ground state of Nd$^{3+}$ was found to be a Kramers doublet with strong Ising anisotropy ($g_\perp $=0) \cite{lhotel2015,xu2015}. Huang et.\ al.\ found that the doublet is a dipolar-octupolar doublet and the symmetry allowed nearest-neighbor exchange interaction takes the form of the XYZ model, a non-Ising interaction which can enhance the spin-flip process \cite{huang2014}. The XYZ model for Nd pyrochlores supports two distinct QSI phases beside the conventional long-range "all-in-all-out" antiferromagnetic order. Even though Nd$_2B_2$O$_7$ ($B=$Zr, Hf, Sn) pyrochlores were reported recently to exhibit the "all-in-all-out" order, strong spin fluctuations can be still expected as suggested by the strongly reduced ordered moment at base temperature \cite{lhotel2015,xu2015,anand2015,bertin2015}. In addition, the $T^3$ dependence of the low temperature magnetic heat capacity of Nd$_2$Sn$_2$O$_7$ and the high susceptibility of the ordered phase of Nd$_2$Zr$_2$O$_7$ also point to the strong fluctuations in the ordered phase \cite{bertin2015,lhotel2015}.    

Very recently, magnetic fragmentation and novel excitations were observed in neutron scattering experiments on a single crystal of Nd$_2$Zr$_2$O$_7$ evidenced by the coexistence of the Bragg peaks from the "all-in-all-out" order and the pinch point pattern characteristic for the magnetic Coulomb phase \cite{petit2016}. It was revealed that the ground state of Nd$_2$Zr$_2$O$_7$ is possibly a magnetic monopole crystal (alternating "3-in-1-out" and "1-in-3-out" spin configurations on the network of tetrahedra) and the fragmentation of the magnetic field of this moment configuration produces the two experimental features \cite{petit2016,brooks2014}. It is very interesting to gain some insight into the spin dynamics of this system.

Recently, the muon spin relaxation ($\mu$SR) technique was used to study the spin dynamics of Nd$_2$Sn$_2$O$_7$ and persistent spin dynamics were found in the ground state along with the static order indicated by the oscillation of the muon polarization \cite{bertin2015}. Here we study Nd$_2$Zr$_2$O$_7$ also using the $\mu$SR technique and find that, in contrast to Nd$_2$Sn$_2$O$_7$, this compound shows strong quantum fluctuations in the ordered state which wash out the normal $\mu$SR signatures for static order. Anomalously slow spin dynamics in the paramagnetic state are also found. Our results support the proposed theoretical analyses mentioned above.

\section{\label{ExpDetails} EXPERIMENTAL DETAILS}

The same Nd$_2$Zr$_2$O$_7$ powder sample was used from Ref. \cite{xu2015} which was synthesized by the conventional solid state reaction method and characterized by synchrotron x-ray diffraction which confirmed a site mixing of less than 0.5~\% and no oxygen defect. The longitudinal muon spin relaxation experiment was performed on the spectrometers LTF (20~mK~-~4.2~K) and GPS (1.5~-~300~K) at the Paul Scherrer Institute (PSI), Switzerland. For a positive muon spin relaxation experiment, the fully polarized muons are implanted into the sample and after a short time (less than $10^{-9}s$) they thermalize without losing their polarization and stop at interstitial crystallographic sites. Then, the muon spins start to precess around the local magnetic field created by the atomic moments. When a muon decays (the life time of the muon is $\tau\approx2.2\times10^{-6}~s$), a positron is emitted with its momentum preferentially along the direction of the muon spin. The number of the emitted positrons is recorded as a function of time by two detectors placed at the forward (F) and backward (B) positions with respect to the initial muon polarization direction. After counting millions of muons, the time evolution of the muon spin polarization (or asymmetry) can be obtained by
\begin{equation}
A(t) = \frac{N_B(t)-\alpha N_F(t)}{N_B(t)+\alpha N_F(t)}=a_0P_z(t)+a_{bg}
\label{eq:princ}
\end{equation}
where $N_B$(t) and $N_F$(t) are the number of counts recorded by the backward, forward detectors, respectively and $\alpha$ is the relative efficiency factor of the two detectors; $a_0$ and $a_{bg}$ are the initial asymmetry and background.

For the experiment on LTF, about 2~g Nd$_2$Zr$_2$O$_7$ powder was mixed with ethanol diluted GE 7031 varnish and attached to a silver plate for a good thermalization. A dilution refrigerator was used which achieved a base temperature of 20~mK and a maximum temperature 4.2~K. The high temperature spectra (1.5~-~300~K) were recorded on GPS with a powder sample of about 2~g wrapped in thin aluminium foil and was cooled by a Helium flow crystat. The data were collected at several temperatures in zero field (ZF) and in several longitudinal fields (LF) (parallel to the initial muon spin polarization) up to 0.5~T and the analysis was done by using the free software package {Musrfit} \cite{musrfit}.

\section{\label{Results_lowT}Results for low temperatures ( T $\leq$ 4.2~K) }
\subsection{\label{ZF}Zero field $\mu$SR measurements}
\begin{figure}
\hspace{-0.8cm}
\includegraphics[width=0.5\linewidth]{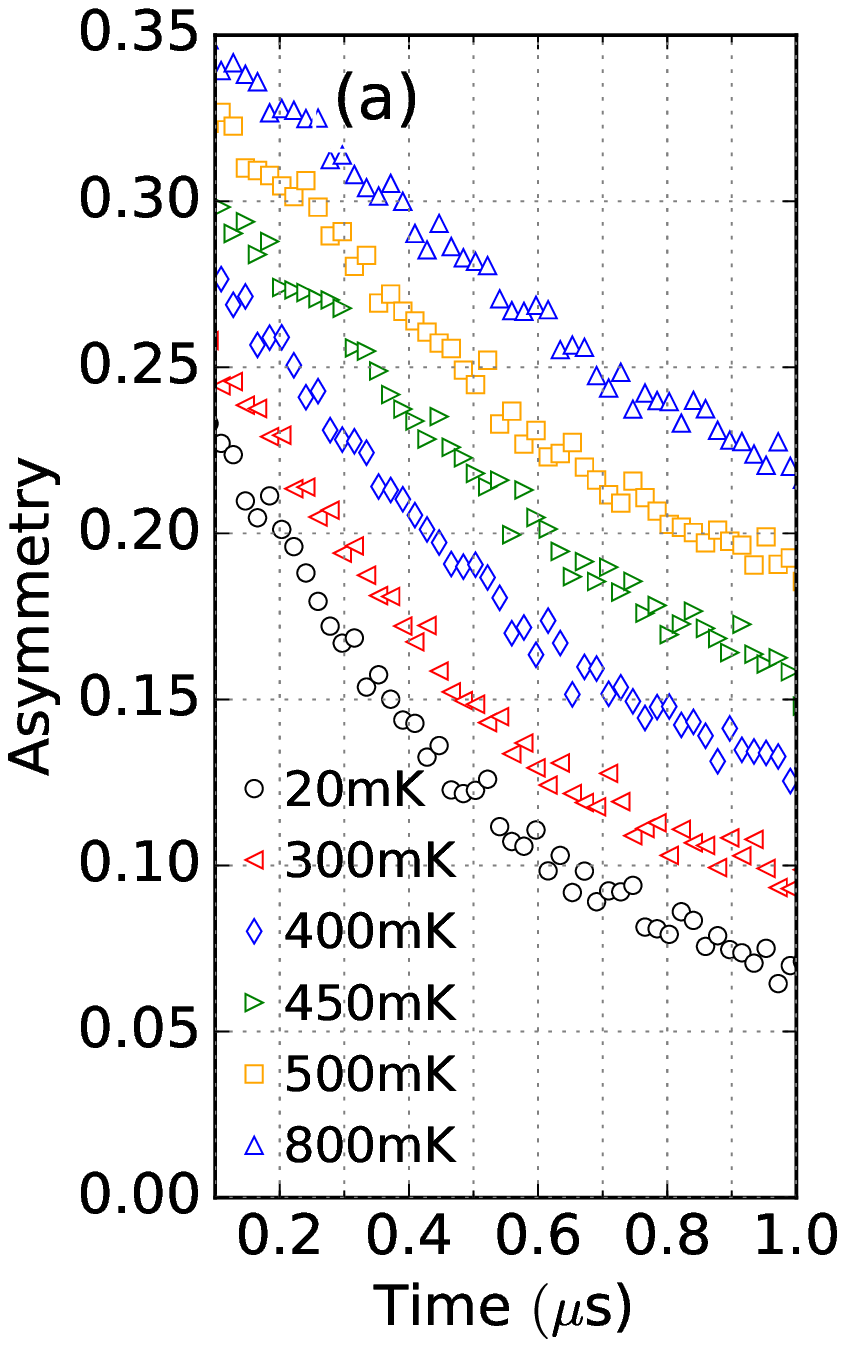}
\includegraphics[width=0.5\linewidth]{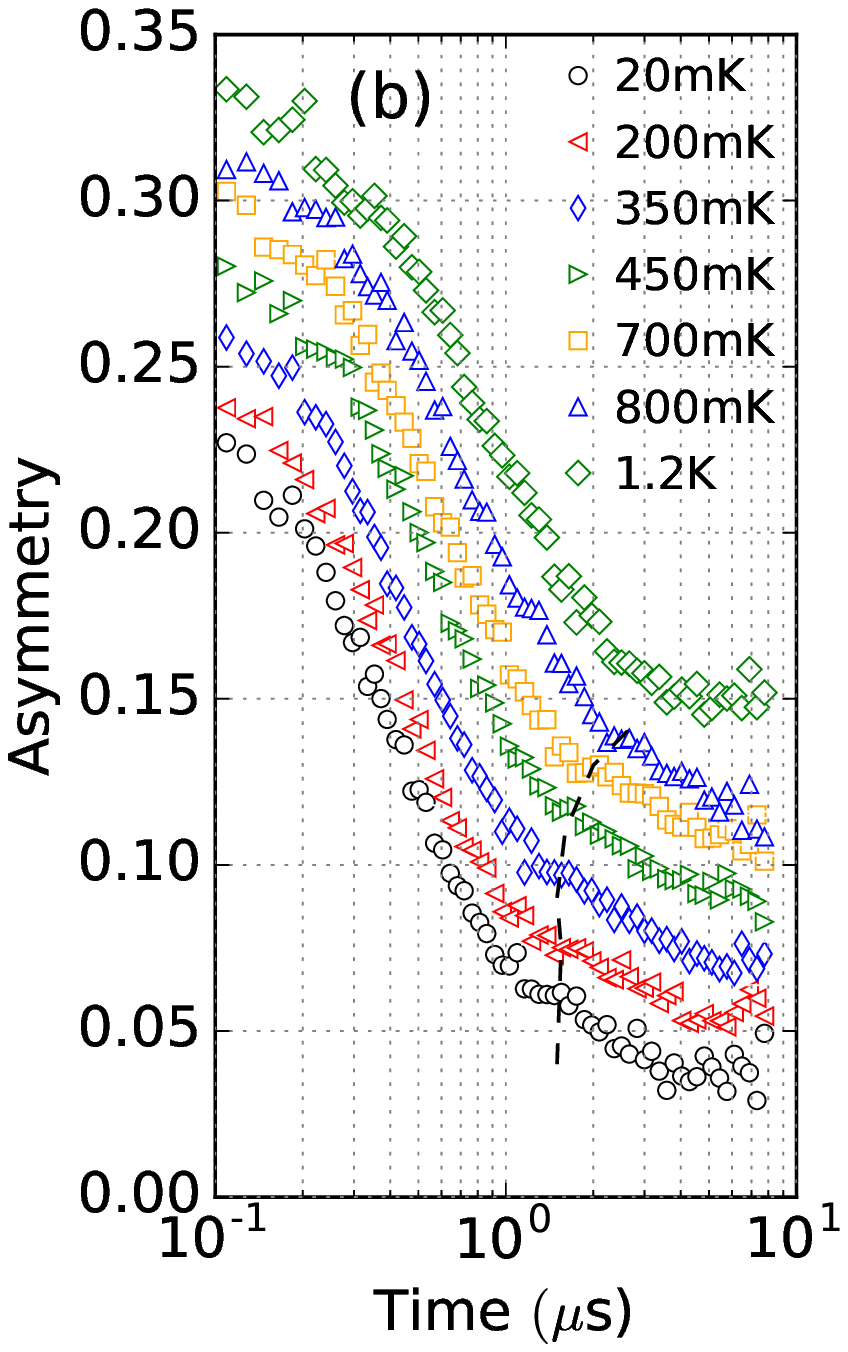}
\hspace{-0.8cm}
\caption{(color online). (a): Early-time weak Gaussian relaxation of the ZF $\mu$SR asymmetry spectra collected at temperatures below 4.2~K. The data are shifted vertically by 0.023 unit successively for a better visualization. (b): Inflection point at 1.5~-~3~$\mu$s and its evolution (dashed line) in the ZF spectra at different temperatures as a function of $\log(Time)$. The data are shifted vertically by 0.017 unit successively for a better visualization.}
\label{fig:ltfzf_gl}
\end{figure}

Figure~\ref{fig:ltfzf_gl} shows representative asymmetry spectra in zero field below 4.2~K. It should be pointed out first that the spectra below the transition temperature ($T_{\rm N}=$~0.4~K) do not show the cosine-function-like oscillations which is a normal signature for static order as found in Nd$_2$Sn$_2$O$_7$ \cite{bertin2015}. In addition, the 1/3$P_z(t=0)$ tail is also not shown which is another indicator for a static internal field in a powder sample \cite{muonbook}. The spectra below and above $T_{\rm N}$ do not show strong temperature dependence. At lower temperatures ($T<0.8$~K), the spectra have a weak Gaussian relaxation at early times (Fig. \ref{fig:ltfzf_gl}(a)) and an inflection point at 1.5~-~3$\mu$s with a tiny local minimum (Fig. \ref{fig:ltfzf_gl}(b)) which are typical for the dynamical Gaussian Kubo-Toyabe function (GKT) at the quasi-static limit (the Gaussian field distribution width and fluctuation rate are comparable and for this case, both are in microsecond range). These two features becomes less pronounced at elevated temperatures such that the spectra become more like an exponential function due to the enhanced thermal spin fluctuations (e.g. the spectra at temperatures above 1.2 K).

\begin{figure}
\hspace{-0.8cm}
\includegraphics[width=3in]{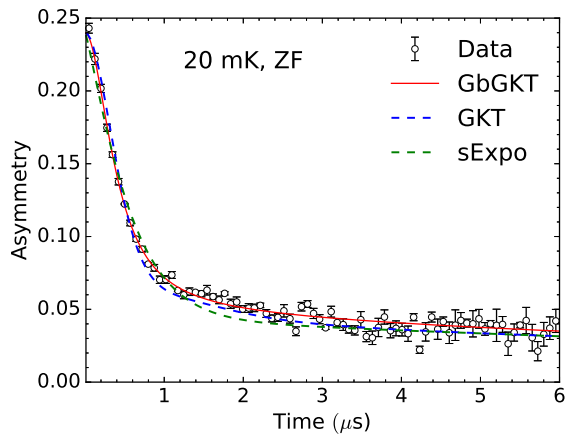}
\hspace{-0.8cm}
\caption{(color online). Zero field asymmetry spectra (open circles) at base temperature fitted with dynamical Gaussian-broadened Gaussian Kubo-Toyabe function (GbGKT) (red solid line), the conventional dynamical Gaussian Kubo-Toyabe function (GKT) (blue dash line) and the stretched exponential function (sExpo) (Eq.~\ref{eq:strech} in Sec.~\ref{Results_highT}) (green dashed line). The corresponding $\chi^2$ for the fitting with these three functions are 1.02, 1.17 and 1.30, respectively.}
\label{fig:20comp}
\end{figure}

\begin{figure}
\hspace{-0.8cm}
\includegraphics[width=3in]{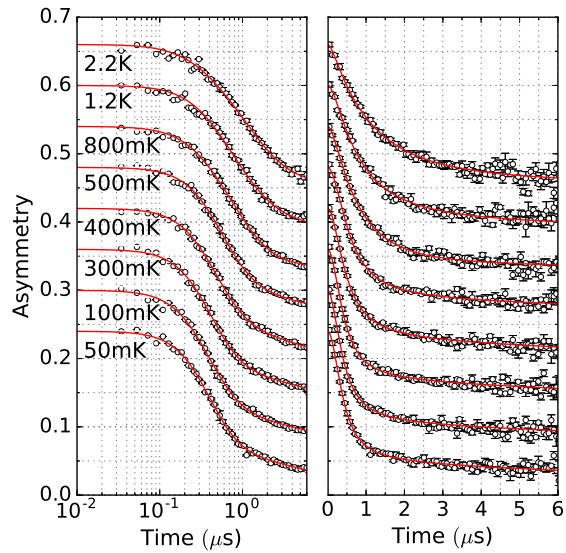}
\hspace{-0.8cm}
\caption{(color online). Logarithmic (left) and linear (right) plots of $\mu$SR asymmetry spectra (open circles) at different temperatures in zero field from LTF instrument and the corresponding fittings (red solid lines) with dynamical GbGKT function. The data are shifted vertically by 0.06 unit successively for a better visualization and the temperature for every spectrum is given by the text nearby.}
\label{fig:ltfzf}
\end{figure}

\begin{figure}
\hspace{-0.8cm}
\includegraphics[width=2.5in]{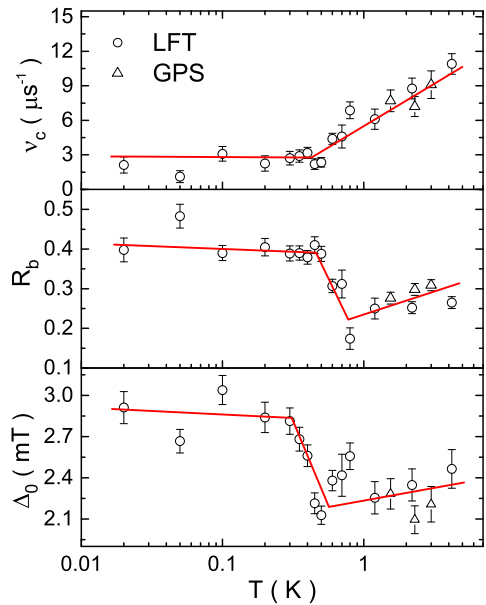}
\hspace{-0.8cm}
\caption{(color online). Temperature dependence of the fitted parameters of the dynamical GbGKT function whose meanings are explained in the main text. The red solid lines are a guide to the eye.}
\label{fig:ltfzf_par}
\end{figure}

The conventional dynamical Gaussian Kubo-Toyabe function (GKT) does not fit the data well because it gives a deeper dip (Fig.~\ref{fig:20comp}) similar to the situation of Yb$_2$Ti$_2$O$_7$ \cite{hodges2002} and Yb$_2$Sn$_2$O$_7$ \cite{yaouanc2013}. The dynamical Gaussian-broadened Gaussian Kubo-Toyabe function (GbGKT) fits the data better which is a phenomenological model accounting for the homogeneous disorder by including a collection of Gaussian field distributions with a Gaussian-function-weighed standard deviation \cite{msrfitmanual, muonbook}. There are three parameters in the model: the fluctuation rate $\nu_c$, the average standard deviation $\Delta_0$ and the standard deviation $\Delta_{\rm G}$ of the Gaussian-weighted standard deviation of the internal field \cite{msrfitmanual}. While for the last one, $R_{\rm b}=\frac{\Delta_{\rm G}}{\Delta_0}$ is usually used instead. The fitted spectra and parameters are shown in Fig.~\ref{fig:ltfzf} and Fig.~\ref{fig:ltfzf_par}, respectively. Note that the fitting reproduces most of the features of the spectra but it does not show a clear inflection point and local minimum. The fitting shows that the fluctuation rate $\nu_c$ decreases with decreasing temperature continuously and becomes almost temperature independent below the ordering temperature, indicating persistent spin dynamics. The parameters $\Delta_0$ and $R_{\rm b}$ characterizing the standard deviation of the field distribution decrease gradually with decreasing temperature and then increase suddenly at the ordering temperature. 

There are mainly two reasons for the absence of oscillations in an ordered state: either there is no local field at the muon site, {i.e.} $B_{\rm loc}$~=~zero or the fluctuations are fast, {i.e.} $\nu_c \gg \gamma _\mu B_{\rm loc}$ ($\gamma_\mu$ is the muon gyromagnetic ratio). The first case can be due to a cancellation of the dipolar field at the muon stopping sites from different ordered moments \cite{muonbook}. Dipolar field cancellation can happen at high symmetry sites in the unit cell for the current highly-symmetric magnetic structure. Considering the sites of oxygen anions near to which the positively charged muon generally stops in oxides, the crystallographic 8$a$ site has a high symmetry. It is at the centre of the Nd$^{3+}$ tetrahedron where all the four spins at the vertices of the tetrahedron point inwards or outwards. At this site the local field is exactly zero. However, a muon normally stops at a distance 0.9~-~1.1~\AA~from the oxygen which is not the centre of the tetrahedron \cite{muonbook}. According to our dipolar field calculation based on the current magnetic structure, there is almost no place surrounding the oxygen ions at the 8$a$ and 48$f$ sites that has zero field \cite{musoftw}. 

In addition, the related compounds Nd$_2$Sn$_2$O$_7$ and Nd$_2$Hf$_2$O$_7$ which have the same ordered structure show oscillations or missing asymmetry (fast oscillations that are not resolved by the instrument), which indicates that there are non-zero static (relative to $\gamma _\mu B_{\rm loc}$) fields at the muon stopping sites in these two compounds \cite{bertin2015,privat2016}. For Nd$_2$Zr$_2$O$_7$, it is reasonable to assume that muons stop at similar positions as in Nd$_2B_2$O$_7$ ($B=$~Sn, Hf) where a non-zero local field should also be present. Moreover, because Nd$_2$Zr$_2$O$_7$ and Nd$_2$Sn$_2$O$_7$ show similar magnitude of ordered moment \cite{xu2015,bertin2015}, a similar oscillation frequency would be expected. While Nd$_2$Sn$_2$O$_7$ shows clear oscillation signal below 0.65~$T_{\rm N}$, Nd$_2$Zr$_2$O$_7$ does not show any sign of oscillation even at the temperature 0.05~$T_{\rm N}$, which indicates much stronger dynamics in Nd$_2$Zr$_2$O$_7$. It should be mentioned that a very broad distribution of static field (compared with the average value of the field) can also wash out the oscillation signal. However, the $\mu$SR spectra for Nd$_2$Sn$_2$O$_7$ indicate that the local field at the muon stopping site can be 127~mT \cite{bertinthesis} and if we assume that Nd$_2$Zr$_2$O$_7$ has a similar static order to Nd$_2$Sn$_2$O$_7$, a similar local field can be expected which is at least an order of magnitude larger than the field distribution width obtained here. The measurement with longitudinal field in Sec.\ref{LF} also excluded this possibility.

Therefore, the absence of the oscillations is attributed to the strong spin fluctuations in the ordered state which is unusual for a system with strong Ising anisotropic moments. This result supports the existence of a significant transverse coupling between the Ising spins possibly induced by the multipole-multipole interactions. Nd$_2$Zr$_2$O$_7$ seems quite different from its counterpart compound Nd$_2B_2$O$_7$ ($B=$~Sn, Hf) as mentioned above, which is not so surprising in the context that sibling pyrochlore compounds could show quite different properties, e.g. Yb and Tb titanates and stannates. According to the microscopic calculation of the psuedo spin exchange interaction in Ref. \cite{onoda2011,rau2015}, the strength ratio of the transverse coupling to the Ising one is strongly dependent on composition of crystal field ground state wave function and thus dependent on the $B$ ions and the crystal field parameters.

\begin{figure}
\hspace{-0.8cm}
\includegraphics[width=3in]{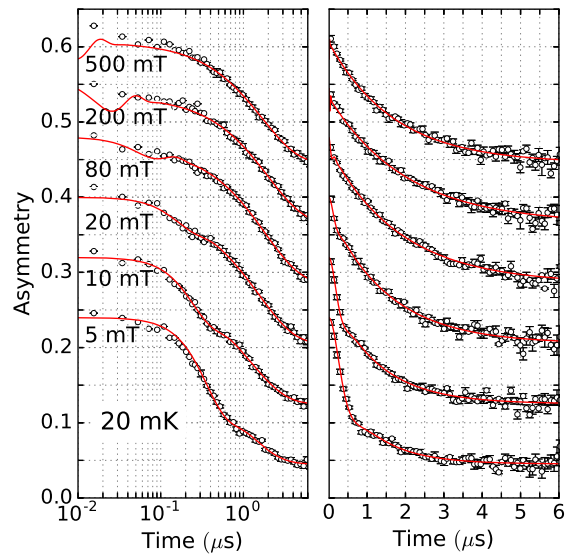}
\hspace{-0.8cm}
\caption{(color online). Logarithmic (left) and linear (right) plots of the spectra (open circles) recorded at base temperature in different longitudinal fields (indicated by the texts nearby) and corresponding fits (red solid lines) with the dynamical GKT function with longitudinal field. The data are shifted vertically by 0.08 unit successively for a better visualization.}
\label{fig:ltf20d}
\end{figure}

\begin{figure}
\hspace{-0.8cm}
\includegraphics[width=2.5in]{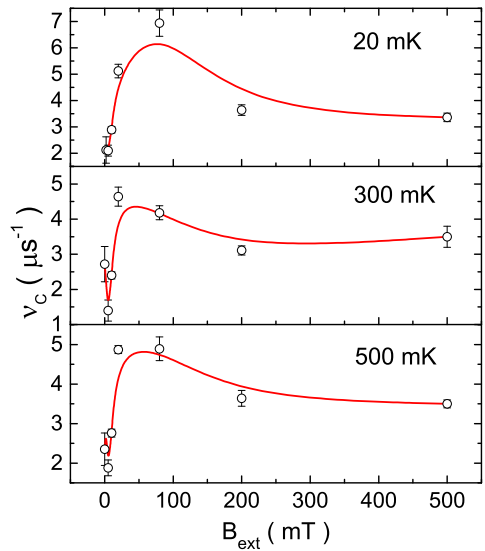}
\hspace{-0.8cm}
\caption{(color online). The fitted fluctuation rate (back circles) in different longitudinal fields $B_{\rm{ext}}$ for temperatures 20~mK, 300~mK and 500~mK. The ZF parameters are also shown which are extracted from the fitting with the dynamical GbGKT function. The red lines are a guide to the eye.}
\label{fig:ltf_drate}
\end{figure}

The obtained spin fluctuation rate is in the range of megahertz and is compatible  with the magnetic structure which has a time scale of larger than $\sim$10$^{-12}$~s as estimated from the energy resolution of the neutron diffractometer \cite{dalm2006,xu2015}. An ordered phase without a spontaneous muon spin precession was also reported for Tb$_2$Sn$_2$O$_7$ \cite{dalm2006,bert2006}, Er$_2$Ti$_2$O$_7$ \cite{lago2005} and Yb$_2$Sn$_2$O$_7$ \cite{yaouanc2013}. However Nd$_2$Zr$_2$O$_7$ is a special case because it is a strong Ising system with additional multipole interactions that introduces the dynamics while for the others a planar component of the spin or low-energy crystal field excitations play vital roles \cite{rev2010,huang2014,mola2007}. In addition, it is very interesting that the obtained fluctuation rate and field distribution width in the ordered and paramagnetic states are comparable with the quantum spin ice materials Yb$_2$Ti$_2$O$_7$ (where magnetic order was found recently) and Yb$_2$Sn$_2$O$_7$ but are much lower than those of the dynamical spin ice Tb$_2$Sn$_2$O$_7$ \cite{dalm2006,bert2006,yaouanc2013,mais2015,hodges2002,gaudet2016}.

It is striking that the spectra in the paramagnetic state are also described by the model at the quasi-static limit rather than the simple exponential or the stretched exponential function which is generally the motion-narrowing limit of the model. This points to anomalously slow paramagnetic spin dynamics which can be associated with the strong spin correlations or short-range order as was also reported for the sibling compound Nd$_2$Sn$_2$O$_7$ and the quantum spin ice materials Yb$_2$Ti$_2$O$_7$, Yb$_2$Sn$_2$O$_7$ \cite{bertin2015,yaouanc2013,hodges2002,mais2015}. This is consistent with the magnetic heat capacity which develops at around 10~K \cite{xu2015}. The slow paramagnetic spin dynamics unexpectedly persists in the ordered state resulting in the absence of the oscillations in the ordered state. 

Recently, a muon induced effect is reported for Pr$_2B_2$O$_7$ ($B=$~Zr, Hf, Sn) which makes the system order because the implanted muon stopping around the 48$f$ site oxygen modifies the local environment anisotropically of the non-Kramers rare earth ion Pr$^{3+}$ and lifts the degeneracy of the crystal field ground state doublet \cite{foronda2015}. This is not expected for the Kramers ion like Nd$^{3+}$ whose Kramers degeneracy is protected by time-reversal symmetry.

\subsection{\label{LF}Longitudinal field $\mu$SR measurements}
The strong fluctuations of the ordered state are also confirmed by measurements in a longitudinal field which is usually used to find out whether the damping of the muon polarization is caused by dephasing due to the distribution of static field or relaxation due to field fluctuations \cite{muonbook}. The asymmetry spectra measured in several fields at 20~mK are shown in Fig.~\ref{fig:ltf20d}. The spectra show clear field-induced oscillations which evolve gradually with increasing field but no sign of relaxation quenching can be seen, which confirms the dynamical properties of the system. 

\begin{figure}
\hspace{-0.8cm}
\includegraphics[width=2.5in]{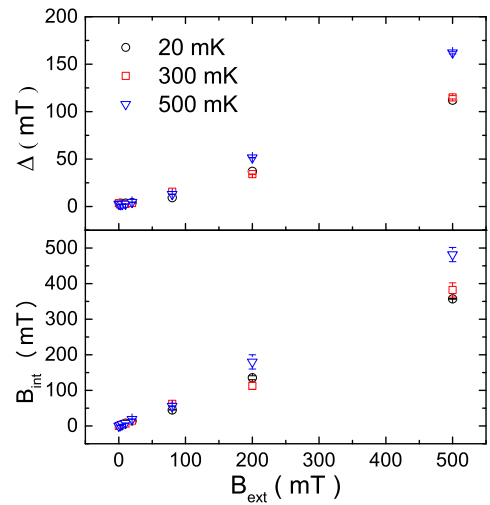}
\hspace{-0.8cm}
\caption{(color online). Longitudinal field $B_{\rm{ext}}$ dependence of the fitted internal field $B_{\rm{int}}$ (lower panel) and the fitted internal field distribution standard deviation $\Delta$ (higher panel) at temperatures 20~mK, 300~mK and 500~mK.}
\label{fig:ltf_ddelta}
\end{figure}

\begin{figure}
\hspace{-0.8cm}
\includegraphics[width=3in]{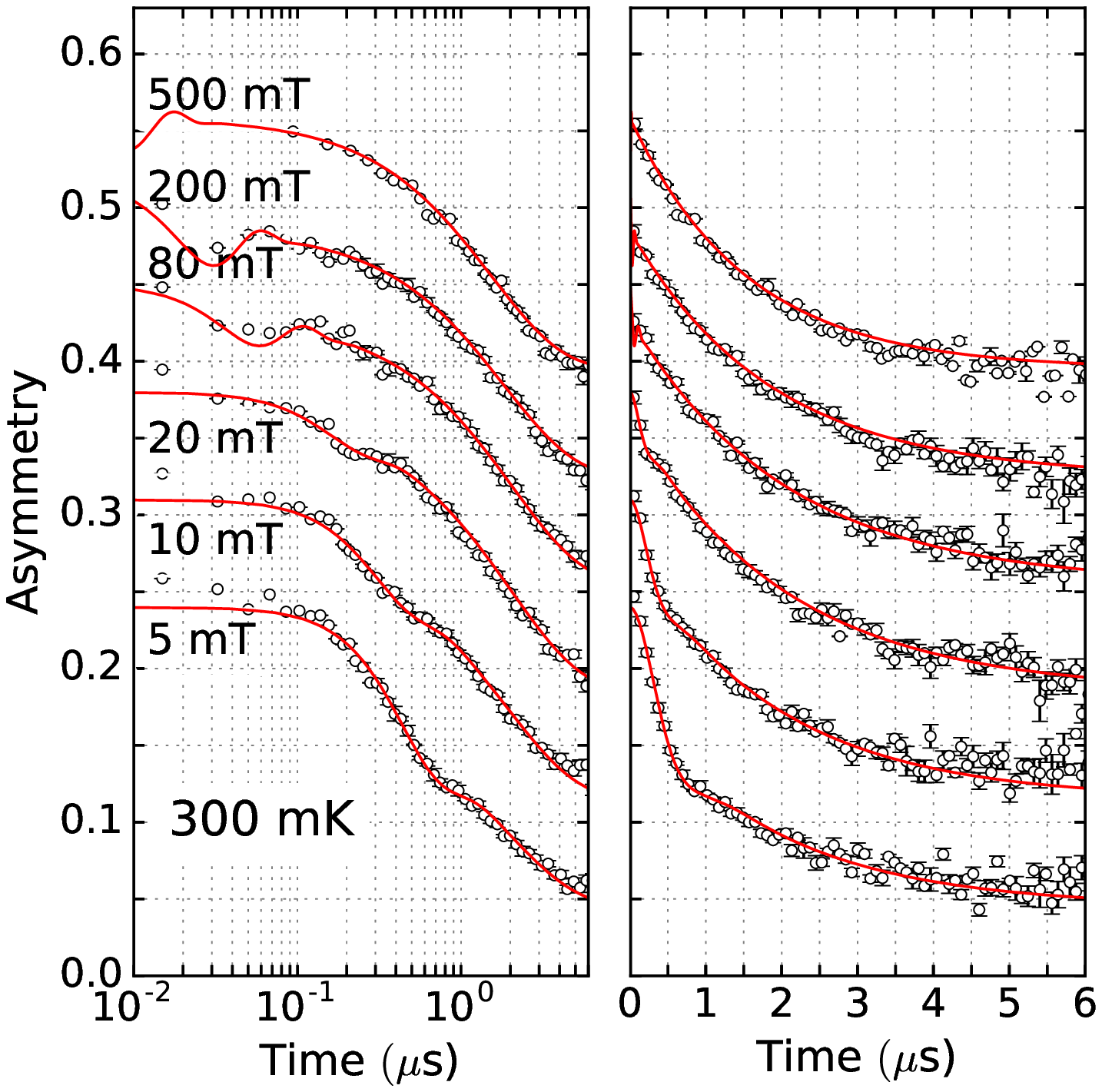}
\hspace{-0.8cm}
\caption{(color online). Logarithmic (left) and linear (right) plots of the spectra (open circles) recorded at 300~mK in different longitudinal fields (indicated by the texts nearby) and corresponding fits (red solid lines) with the dynamical GKT function with longitudinal field. The data are shifted vertically by 0.07 unit successively for a better visualization.}
\label{fig:ltf300d}
\end{figure}

\begin{figure}
\hspace{-0.8cm}
\includegraphics[width=3in]{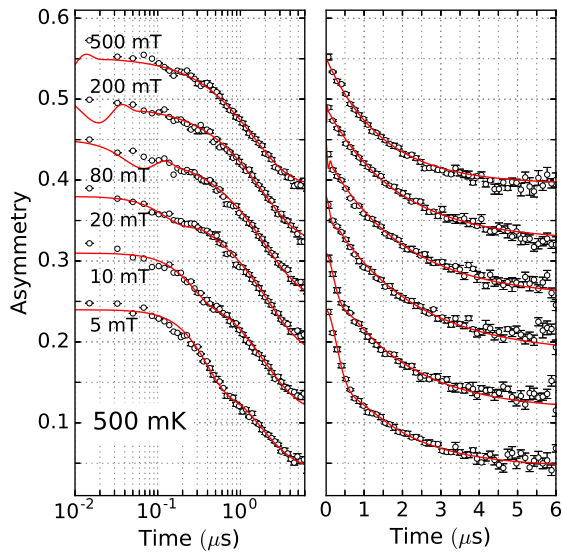}
\hspace{-0.8cm}
\caption{(color online). Logarithmic (left) and linear (right) plots of the spectra (open circles) recorded at 500~mK in different longitudinal fields (indicated by the texts nearby) and corresponding fits (red solid lines) with the dynamical GKT function with longitudinal field. The data are shifted vertically by 0.07 unit successively for a better visualization.}
\label{fig:ltf500d}
\end{figure}

The dynamical Gaussian Kubo-Toyabe function with longitudinal field gives a good description of the data. The fitted curves for the 20~mK data in different external fields $B_{\rm ext}$ are shown in Fig.~\ref{fig:ltf20d} and the field dependence of the parameters are shown in Fig.~\ref{fig:ltf_drate} and Fig.~\ref{fig:ltf_ddelta}. As the applied field increases, the fluctuation rate $\nu_c$ increases and shows a maximum around 80~mT, which is not a simple field quenching effect. We tentatively ascribe this peak in the fluctuation rate to the field-induced metamagnetic transition from the antiferromagnetic spin configuration to a ferromagnetic one as observed in the single crystal magnetization measurement (the transition fields for the three main cubic directions are around 100~mT at 90~mK) \cite{lhotel2015}. The same model also fits the data at 300~mK (below $T_{\rm N}$) and 500~mK (above $T_{\rm N}$) as shown in Fig.~\ref{fig:ltf300d} and Fig.~\ref{fig:ltf500d}. The corresponding parameters are shown in Fig.~\ref{fig:ltf_drate} and Fig.~\ref{fig:ltf_ddelta}. A similar field induced peak in the fluctuation rate for 300~mK is also observed but at a lower field, similar to the magnetization data which shows a decreased metamagnetic transition field at a higher temperature \cite{lhotel2015}. However, at 500~mK in the paramagnetic state, the peak is still present, which might be due to the break down of the short-range antiferromagnetic correlations induced by the field (discussed more in the next section).

In addition, the fitted internal field $B_{\rm int}$ and field standard deviation $\Delta$ increase with increasing the external field almost linearly (Fig.~\ref{fig:ltf_ddelta}) similar with Yb$_2$Ti$_2$O$_7$ and Yb$_2$Sn$_2$O$_7$ \cite{mais2015} but the fitted internal field is much smaller than the applied field for the two temperatures below $T_{\rm N}$. If we associate $\mid B_{\rm int}-B_{\rm ext}\mid $ with the demagnetizing field, this agrees with magnetization data which shows a higher magnetization at lower temperatures \cite{lhotel2015}. 

\begin{figure}
\includegraphics[width=3in, keepaspectratio]{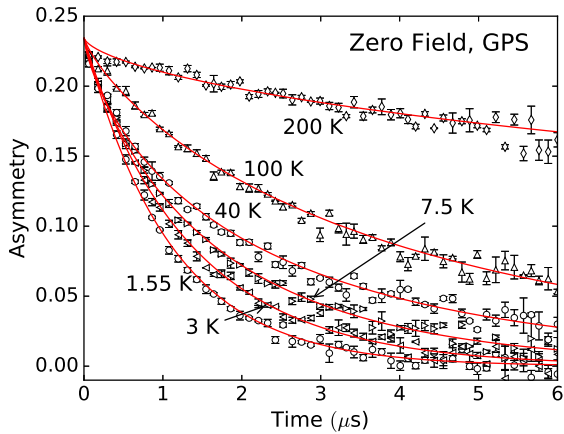}
\caption {(color online). Selected spectra (open circles) recorded on the GPS instrument in zero field at different temperatures between 1.55 and 300~K. The temperature for each spectrum is indicated by the text nearby. The red lines show the fitting with stretched exponential function.}
\label{fig:gpszf}
\end{figure}

\section{\label{Results_highT}Results for High Temperatures ( T $\geq$ 4.2~K) }

\begin{figure}
\includegraphics[width=2.5in, keepaspectratio]{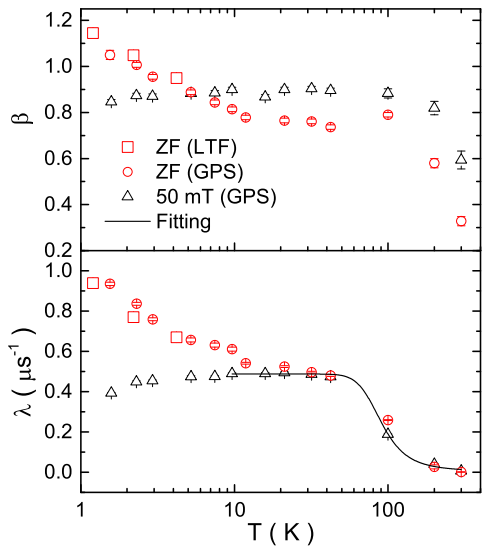}
\caption {(color online). Temperature dependence of the fitted exponent $\beta$ (the upper panel) and relaxation rate $\lambda$ (the lower panel). The data from LTF in zero field and the data from GPS in zero and 50~mT fields are shown. The black line is the fitting of the temperature dependence of the relaxation rate due to the Orbach process for the 50~mT data.}
\label{fig:gpspar}
\end{figure}

The data at temperatures between 1.55~K and 300~K in zero field from the GPS instrument are shown in Fig.~\ref{fig:gpszf}. The initial Gaussian-type damping and the tiny local minimum shown in the low-temperature data become very weak or disappear. The spectra are well fitted with the stretched exponential function
\begin{equation}
P_Z(t) = a_0\exp[-(\lambda t)^\beta ]
\label{eq:strech}
\end{equation}
which is commonly used for a system with a continuous distribution of relaxation processes \cite{muonbook}. Note that the spectra in the temperature range 1.2~-~4.2~K initially fitted with the dynamical GbGKT function in Sec.\ref{ZF} are also described well by the stretched exponential function ($\beta>$~1 for the lower temperatures) and are also shown in Fig.~\ref{fig:gpszf}. The fitted relaxation rate $\lambda$ and exponent $\beta$ are shown in Fig.~\ref{fig:gpspar} as a function of temperature. In zero field, the relaxation rate $\lambda$ increases with decreasing temperature indicating slowing down of the spin dynamics, which is consistent with the results fitted with the dynamical GbGKT function (Fig. \ref{fig:ltfzf_par}). At the same time, the exponent $\beta$ increases and is larger than one below 2~K showing the evolution of the spectra, which further indicates the slowing down of the system. The increase in the relaxation rate or the slowing down of the spin fluctuations below 20~K is due to the approach to the magnetic ordering as observed in the specific heat measurement \cite{xu2015}. This increase of $\lambda$ is largely suppressed by the 50~mT longitudinal field, as was also found in Nd$_2$Sn$_2$O$_7$ \cite{bertin2015}, which maybe caused by the competition between the antiferromagnetic exchange interaction and Zeeman interaction (explained below).

The inflection point of $\lambda$ at $\sim$80~K could be induced by Orbach process: here the magnetic moment of Nd$^{3+}$ relaxes through a real two phonon process with an excited crystal-field state as a intermediate state. The temperature dependence of the relaxation due to the Orbach process can be described by \cite{dalmas2003}
\begin{equation}
\lambda^{-1} = \lambda_0^{-1} + \eta\exp[-{\rm \Delta_{CEF}}/(k_{\rm B} T)]
\label{eq:orbach}
\end{equation}
where $\lambda_0^{-1}$ is the saturation value at low temperatures, $\eta$ reflects the strength of the spin-lattice interaction and $\Delta_{\rm CEF}$ is energy gap between the crystal field ground state and the excited state that mediates the spin relaxation process. The fitting of the extracted relaxation rate $\lambda$ in the field of 50~mT yields $\Delta_{\rm CEF}=~$40(4)~meV (with $\lambda_0^{-1}=2.0(1)~\mu\rm{s}$, $\eta=335(30)~\mu{\rm s}$) which is close to the energy of the second and third excited crystal field levels (34.4 and 35.7 meV) determined by the inelastic neutron scattering experiments \cite{xu2015}. Because the two crystal field levels are quite close to each other, fitting with the current data does not allow to distinguish which one participates in the Orbach process. A qualitatively similar result was also found for Nd$_2$Sn$_2$O$_7$ \cite{bertinthesis}.

\begin{figure}
\includegraphics[width=3in, keepaspectratio]{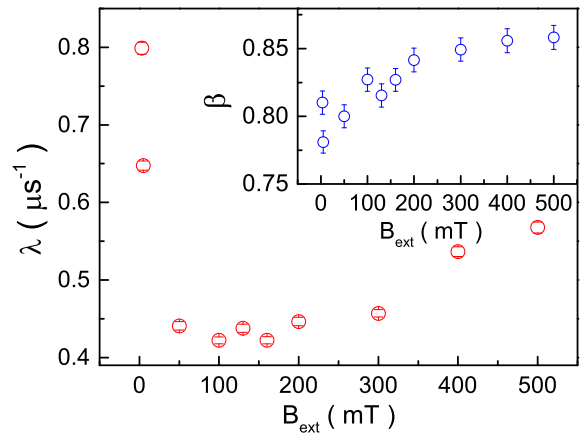}
\caption {(color online). Field dependence of the fitted relaxation rate $\lambda$ and exponent $\beta$ (inset) of the GPS data at 1.55~K.}
\label{fig:gps_dpar}
\end{figure}

As mentioned above, a 50~mT field largely suppresses the increase of $\lambda$ at 1.5~$\sim$~20~K. At first, it seems that it can be explained using the Redfield formula \cite{muonbook},
\begin{equation}
\lambda =\frac{2\gamma_{\mu}^2\Delta^2\nu_c}{(\gamma_{\mu}B_{\rm ext})^2+\nu_c^2},
\label{eq:redfield}
\end{equation}
which shows that assuming the applied field has no influence on the system and in the fast fluctuation limit, the longitudinal field dependence of the relaxation rate can be described by a Lorentzian function with the half width at half maximum to be $\nu_c/\gamma_{\mu}$. Accordingly, for a system with a very small fluctuation rate, $\lambda$ can be reduced sharply to nearly zero by the field. Fig.~\ref{fig:ltfzf_par} shows that the fluctuation rate in zero field of Nd$_2$Zr$_2$O$_7$ is of the order of magnitude of 10~MHz corresponding to $\sim$12~mT which is much small that 50~mT. Therefore, it seems that the strong reduction in $\lambda$ by the 50~mT field is probably due to relatively low paramagnetic fluctuation rate of the system. However, Fig.~\ref{fig:gps_dpar} shows that with the application of a higher field (below 200~mT), $\lambda$ does not fall to be zero but keeps constant at $\sim$0.43 which is almost the saturation relaxation rate due to the Orbach process at low temperature. Hence the Redfield formula may be not applicable to this situation. On the other hand, Fig.~\ref{fig:ltf_drate} shows that the fluctuation rate of the paramagnetic state (at 500~mK) can be largely increased by a low external field, which clearly shows that the assumption for the Redfield formula is not satisfied in Nd$_2$Zr$_2$O$_7$. As a result, we think that the increase of $\lambda$ in zero field below 20~K is due to the establishment short-range antiferromagnetic spin correlations which slows down the system. And the applied field competes with antiferromagnetic exchange interaction or breaks the short-range antiferromagnetic correlations resulting the increase of the fluctuation rate and thus $\lambda$ recovers to be the single ion relaxation rate due to the Orbach process. In addition, the high field ($>$~200~mT) increase of $\lambda$ shown in Fig.~\ref{fig:gps_dpar} is possibly due to a crystal field effect.

\section{\label{Conclusion} Discussion and Conclusions}

We have presented a careful $\mu$SR study on the spin dynamics of the dipolar-octuplar psuedospin-$\frac{1}{2}$ system Nd$_2$Zr$_2$O$_7$ which has a long-range "all-in-all-out" order ($T_{\rm N}$=0.4~K) according to the previous neutron diffraction experiments. We found that the spin dynamics slow down with decreasing temperature and become almost temperature independent below the ordering temperature indicating persistent spin dynamics. At 20~mK, far below $T_{\rm N}$, no oscillation or missing asymmetry is observed in the data, which points to strong spin fluctuations in the ordered state. This result confirms the existence of significant transverse terms in the spin Hamiltonian resulting from the multipole interactions which enables the spin dynamics in this system of Ising spins. The data in longitudinal fields also support the dynamical nature of the ground state and suggest a field-induced metamagnetic transition which agrees with the reported magnetization data of a single crystal \cite{lhotel2015}.

The strong spin fluctuations are consistent the scenario of magnetic moment fragmentation which was recently introduced by the theorists and observed in neutron scattering experiments \cite{brooks2014, petit2016}. It has been shown by the neutron data that the ground state of Nd$_2$Zr$_2$O$_7$ is possibly a fluctuating Coulomb phase superimposed on the monopole ordered phase where strong spin fluctuations are expected. In addition, spin excitations in Nd$_2$Zr$_2$O$_7$ were observed at very low energies which may also be the origin of the strong spin fluctuations \cite{petit2016}.

Anomalously slow paramagnetic dynamics were also found for Nd$_2$Zr$_2$O$_7$ related to the short-range order, which has been reported for many frustrated magnets \cite{bertin2015}. Recently, emergent spin clusters, for instance, unidimensional spin loops, were proposed as their origin \cite{yao2015}. The spin loop excitations in the ground state might also lead to the observed persistent spin dynamics below the transition temperature \cite{yao2015}.

\acknowledgements
We thank A. T. M. N. Islam for his help in sample preparation, and V. K. Anand and S. Chillal for helpful discussions.  We acknowledge Helmholtz Gemeinschaft for funding via the Helmholtz Virtual Institute (Project No. VH-VI-521).


\begin{thebibliography}{99}

\bibitem{book}
C. Lacroix, P. Mendels, and F. Mila, \emph{Introduction to Frustrated Magnetism: Materials, Experiments, Theory} (Springer Science \& Business Media, New York, 2011, Vol. 164).

\bibitem{rev2010}
J. S. Gardner, M. J. P. Gingras, and J. E. Greedan, {\it Magnetic pyrochlore oxides}, Rev. Mod. Phys. {\bf 82}, 53 (2010).

\bibitem{castelnovo2008}
C. Castelnovo, R. Moessner, and S. L. Sondhi, {\it Magnetic monopoles in spin ice}, Nature {\bf 451}, 42  (2008).

\bibitem{morris2009}
D. J. P. Morris, D. A. Tennant, S. A. Grigera, B. Klemke, C. Castelnovo,  R. Moessner, C. Czternasty, M. Meissner, K. C. Rule, J.-U. Hoffmann, K. Kiefer, S. Gerischer, D. Slobinsky, and R. S. Perry, {\it Dirac strings and magnetic monopoles in the spin ice Dy$_2$Ti$_2$O$_7$}, Science {\bf 326}, 411 (2009).

\bibitem{fennell2009}
T. Fennell, P. P. Deen, A. R. Wildes, K. Schmalzl, D. Prabhakaran, A. T. Boothroyd, R. J. Aldus, D. F. McMorrow, and S. T. Bramwell, {\it Magnetic coulomb phase in the spin ice Ho$_2$Ti$_2$O$_7$}, Science {\bf 326}, 415  (2009).

\bibitem{lago2005}
J. Lago, T. Lancaster, S. J. Blundell, S. T. Bramwell, F. L. Pratt, M. Shirai, and C. Baines, {\it Magnetic ordering and dynamics in the XY pyrochlore antiferromagnet: a muon-spin relaxation study of Er$_2$Ti$_2$O$_7$ and Er$_2$Sn$_2$O$_7$}, J. Phys.: Condens. Matter. {\bf 17}, 979 (2005).

\bibitem{dalm2006}
P. Dalmas de R\'eotier, A. Yaouanc, L. Keller, A. Cervellino, B. Roessli, 
C. Baines, A. Forget, C. Vaju, P. C. M. Gubbens, A. Amato, and P. J. C. King, {\it Spin dynamics and magnetic order in magnetically frustrated Tb$_2$Sn$_2$O$_7$}, Phys. Rev. Lett. {\bf 96}, 127202  (2006).

\bibitem{bert2006}
F. Bert, P. Mendels, A. Olariu, N. Blanchard, G. Collin, A. Amato, C. Baines, and A. D. Hillier, {\it Direct evidence for a dynamical ground state in the highly frustrated Tb$_2$Sn$_2$O$_7$ pyrochlore}, Phys. Rev. Lett. {\bf 97}, 117203  (2006).

\bibitem{yaouanc2013}
A. Yaouanc, P. Dalmas de R\'eotier, P. Bonville, J. A. Hodges, V. Glazkov, L. Keller, V. Sikolenko, M. Bartkowiak, A. Amato, C. Baines, P. J. C. King, P. C. M. Gubbens, and A. Forget, {\it Dynamical splayed ferromagnetic ground state in the quantum spin ice Yb$_2$Sn$_2$O$_7$}, Phys. Rev. Lett. {\bf 110}, 127207  (2013).

\bibitem{hodges2002}
J. A. Hodges, P. Bonville, A. Forget, A. Yaouanc, P. Dalmas de R\'eotier, G. Andr\'e, G. M. Rams, K. Kr\'olas, C. Ritter, P. C. M. Gubbens, C. T. Kaiser, P. J. C. King, and C. Baines, {\it First-order transition in the spin dynamics of geometrically frustrated Yb$_2$Ti$_2$O$_7$}, Phys. Rev. Lett. {\bf 88}, 077204  (2002).

\bibitem{mais2015}
A. Maisuradze, P. Dalmas de R\'eotier, A. Yaouanc, A. Forget, C. Baines, and P. J. C. King, {\it Anomalously slow spin dynamics and short-range correlations in the quantum spin ice systems Yb$_2$Ti$_2$O$_7$ and Yb$_2$Sn$_2$O$_7$}, Phys. Rev. B {\bf 92}, 094424  (2015).

\bibitem{rau2015}
J. G. Rau and M. J. P. Gingras, {\it Magnitude of quantum effects in classical spin ices}, Phys. Rev. B {\bf 92}, 144417  (2015).

\bibitem{hermele2004}
M. Hermele, M. P. A. Fisher, and L. Balents, {\it Pyrochlore photons: The $U(1)$ spin liquid in a $S=\frac{1}{2}$ three-dimensional frustrated magnet}, Phys. Rev. B {\bf 69}, 064404  (2004).

\bibitem{onoda2010}
S. Onoda and Y. Tanaka, {\it Quantum melting of spin ice: emergent cooperative quadrupole and chirality}, Phys. Rev. Lett. {\bf 105}, 047201  (2010).

\bibitem{onoda2011}
S. Onoda and Y. Tanaka, {\it Quantum fluctuations in the effective pseudospin-1/2 model for magnetic pyrochlore oxides}, Phys. Rev. B {\bf 83}, 094411  (2011).

\bibitem{ross2011}
K. A. Ross, L. Savary, B. D. Gaulin, and L. Balents, {\it Quantum excitations in quantum spin ice}, Phys. Rev. X {\bf 1}, 021002  (2011).

\bibitem{savary2012}
L. Savary and L. Balents, {\it Coulombic quantum liquids in spin-$1/2$ pyrochlores}, Phys. Rev. Lett. {\bf 108}, 037202  (2012).

\bibitem{kimura2013}
K. Kimura, S. Nakatsuji, J. -J. Wen, C. Broholm, M. B. Stone, E. Nishibori, and H. Sawa, {\it Quantum fluctuations in spin-ice-like Pr$_2$Zr$_2$O$_7$}, Nat. Commun. {\bf 4}, 1934 (2013).

\bibitem{lhotel2015}
E. Lhotel, S. Petit, S. Guitteny, O. Florea, M. Ciomaga Hatnean, C. Colin, E. Ressouche, M. R. Lees, and G. Balakrishnan, {\it Fluctuations and all-in--all-out ordering in dipole-octupole Nd$_2$Zr$_2$O$_7$}, Phys. Rev. Lett. {\bf 115}, 197202 (2015).

\bibitem{xu2015}
J. Xu, V. K. Anand, A. K. Bera, M. Frontzek, D. L. Abernathy, N. Casati, K. Siemensmeyer, and B. Lake, {\it Magnetic structure and crystal-field states of the pyrochlore antiferromagnet Nd$_2$Zr$_2$O$_7$}, Phys. Rev. B {\bf 92}, 224430  (2015).

\bibitem{huang2014}
Y.-P. Huang, G. Chen, and M. Hermele, {\it Quantum spin ices and topological phases from dipolar-octupolar doublets on the pyrochlore lattice}, Phys. Rev. Lett. {\bf 112}, 167203  (2014).

\bibitem{anand2015}
V. K. Anand, A. K. Bera, J. Xu, T. Herrmannsd\"{o}rfer, C. Ritter and B. Lake, {\it Observation of long range magnetic ordering in frustrated pyrohafnate Nd$_2$Hf$_2$O$_7$: A neutron diffraction study} Phys. Rev. B {\bf 92}, 184418 (2015).

\bibitem{bertin2015} 
A. Bertin, P. Dalmas de R\'eotier, B. F{\aa}k, C. Marin, A. Yaouanc, A. Forget, D. Sheptyakov, B. Frick, C. Ritter, A. Amato, C. Baines, and P. J. C. King, {\it Nd$_2$Sn$_2$O$_7$: an all-in-all-out pyrochlore magnet with no divergence-free field and anomalously slow paramagnetic spin dynamics}, Phys. Rev. B {\bf 92}, 144423  (2015).

\bibitem{petit2016}
S. Petit, E. Lhotel, B. Canals, M. Ciomaga Hatnean, J. Ollivier, H. Mutka, E. Ressouche, A. R. Wildes, M. R. Lees, and G. Balakrishnan, {\it Observation of magnetic fragmentation in spin ice}, Nat. Phys. {\bf 12}, 746 (2016).

\bibitem{brooks2014}
M. E. Brooks-Bartlett, S. T. Banks, L. D. C. Jaubert, A. Harman-Clarke, and P. C. W. Holdsworth, {\it Magnetic-moment fragmentation and monopole crystallization}, Phys. Rev. X {\bf 4}, 011007 (2014).

\bibitem{musrfit} 
MUSRFIT, a programme developed by Andreas Suter and Bastian Wojek in the Paul Scherrer Institute.

\bibitem{muonbook}
A. Yaouanc and P. Dalmas de R\'eotier, {\it Muon spin rotation, relaxation, and resonance: applications to condensed matter}, Oxford University Press, 2011.

\bibitem{msrfitmanual}
Musrfit User Manual. Note that in Ref.\cite{muonbook} $\Delta_{\rm eff}^2=(1+R_b^2)\Delta_0^2$ is used in the equation instead of $\Delta_0$. 

\bibitem{musoftw}
M$\mu$Calc, a software for calculating fields inside crystals developed by A. J. Steele; {\it Quantum magnetism probed with muon-spin relaxation}, Oxford University, PhD thesis.

\bibitem{privat2016}
V. K. Anand, et. al., unpublished.

\bibitem{bertinthesis}
Alexandre Bertin. Geometrical frustration and quantum origin of spin dynamics,   Universit\'e Grenoble Alpes, PhD thesis.

\bibitem{mola2007}
H. R. Molavian, M. J. P. Gingras, and B. Canals, {\it Dynamically induced frustration as a route to a quantum spin ice state in Tb$_2$Ti$_2$O$_7$ via virtual crystal field excitations and quantum many-body effects}, Phys. Rev. Lett. {\bf 98}, 157204  (2007).

\bibitem{gaudet2016}
J. Gaudet, K. A. Ross, E. Kermarrec, N. P. Butch, G. Ehlers, H. A. Dabkowska, and B. D. Gaulin, {\it Gapless quantum excitations from an icelike splayed ferromagnetic ground state in stoichiometric Yb$_2$Ti$_2$O$_7$}, Phys. Rev. B {\bf 93}, 064406  (2016).

\bibitem{foronda2015}
F. R. Foronda, F. Lang, J. S. M\"oller, T. Lancaster, A. T. Boothroyd, F. L. Pratt, S. R. Giblin, D. Prabhakaran, and S. J. Blundell, {\it Anisotropic local modification of crystal field levels in Pr-based pyrochlores: a muon-induced effect modeled using density functional theory}, Phys. Rev. Lett. {\bf 114}, 017602  (2015).

\bibitem{dalmas2003}
P. Dalmas de R\'eotier, A. Yaouanc, P. C. M. Gubbens, C. T. Kaiser, C. Baines, and P. J. C. King, {\it Absence of magnetic order in Yb$_3$Ga$_5$O$_{12}$: relation between phase transition and entropy in geometrically frustrated materials}, Phys. Rev. Lett. {\bf 91}, 167201  (2003).

\bibitem{yao2015}
A. Yaouanc, P. Dalmas de R\'eotier, A. Bertin, C. Marin, E. Lhotel, A. Amato, and C. Baines, {\it Evidence for unidimensional low-energy excitations as the origin of persistent spin dynamics in geometrically frustrated magnets}, Phys. Rev. B {\bf 91}, 104427  (2015).

\end{thebibliography}
\end{document}